\documentclass[12pt,a4paper]{article}
\usepackage{amssymb,amsmath,amscd,epsfig}
\usepackage[dvips]{color}
\newcommand{\be}{\begin{eqnarray}}
\newcommand{\ee}{\end{eqnarray}}
\newcommand{\bi}{\bibitem}

\newcommand{\rar}{\rightarrow}
\newcommand{\lrar}{\leftrightarrow}

\newcommand{\tcred}{\textcolor{red}}
\newcommand{\vs}{\vspace}
\newcommand{\nin}{\noindent}
\newcommand{\rv}{\rho_{vac}}
\newcommand{\rc}{\rho_{c}}
\definecolor{gold}{rgb}{0.89,0.78,0}
\definecolor{grn05}{rgb}{0,0.5,0}

\begin{document}

\title{COSMOLOGY AND NEW PHYSICS}
\author{
A.D. Dolgov\\
{\it ITEP, 117218, Moscow, Russia}\\
{\it INFN, Ferrara 40100, Italy}
}
\date{}
\maketitle

\begin{abstract}

A comparison of the standard models in particle physics and in cosmology 
demonstrates that they are not compatible, though both are well established. 
Basics of modern cosmology are briefly reviewed. It is argued that the 
measurements of the main cosmological parameters are achieved through many 
independent physical phenomena and this minimizes possible interpretation 
errors. It is shown that astronomy demands new physics beyond the frameworks of 
the (minimal) standard model in particle physics. More revolutionary modifications 
of the basic principles of the theory are also discussed.

\end{abstract}

\section{ Introduction \label{s-intro}}

Particle physics celebrates an excellent agreement of the
Minimal Standard Model (MSM) with experiment, except probably for
neutrino oscillations. On the other hand, cosmology also demonstrates
a good agreement of astronomical observations with the standard
cosmological model (SCM). So far so good, but it seems that MSM and SCM
are not compatible and an explanation of the observed features of 
the universe is impossible without new physical phenomena. 
Astronomical observations have already led and will certainly lead in 
the near future to astonishing discoveries which may shatter 
cornerstones of contemporary physics and
modify our understanding of basic principles.
\vs{0.3cm}\\
The notion of new physics includes quite different levels of novelty:\\
1. New objects and/or interactions.\\
2. Breaking of established rules or conservation laws.\\
3. New principles.\\
This list can possibly be extended.
\vs{0.3cm}\\
I. Some new physics is quite {\it natural} to expect. For example almost
inevitable are:\\
1. New fields or/and particles: stable or quasi-stable, heavy or light.\\
2. Breaking of charges/quantum numbers:\\
a) electric, (impossible? or at least nontrivial with higher dimensions);\\
b) baryonic (practically certain, cosmologically discovered);\\
c) total leptonic (expected);\\
d) leptonic family(discovered in neutrino oscillations!).\\
Less probable:\\
3. Topological or non-topological solitons.\\
4. New types of interactions, especially new long range forces, and
in particular, modified gravity at large distances.\\
5. Higher dimensions. It is unclear if astronomy is more sensitive to them or
high energy physics. If they are small, of microscopic size then chances to ``feel''
them are better in high energy collisions. In the case of large higher dimensions 
astronomy may successfully compete with high energy physics. 
\vs{0.3cm}\\
II. {\it Unnatural new physics} includes\\
1. Breaking of Lorentz-invariance.\\
2. Violation of CPT.\\
3. Breaking of the spin-statistics relation.\\
4. Breaking of unitarity and quantum coherence.\\
5. Violation of energy conservation.\\
6. Violation of causality and possible existence of time-machine.\\
7. Breaking of least action principle, and of Hamilton and Lagrange
dynamics.
\vs{0.3cm}\\
III. {\it Unexpected new physics} - anything which is not in the list
above and  {\it a priori } will never be there.
\vs{0.1cm}\\

The content of these lectures is the following. In the next section the
standard cosmological model is described. In sec.~\ref{s-univ-today}
astronomical data characterizing the present day universe are presented.
In sec.~\ref{s-infl} the necessity of inflation is advocated and
basis features of inflationary cosmology are described. 
Section~\ref{s-baryo} is devoted to cosmological baryogenesis and in
related sec.~\ref{s-cp} cosmological mechanisms of CP-violation are
discussed. In sec.~\ref{s-dark} the problems of vacuum and dark
energies are considered. Basic features of primordial nucleosynthesis 
are briefly presented in sec.~\ref{s-bbn}. Formation of astronomical
large scale structures is discussed in sec.~\ref{s-lss}. 
In sec.~\ref{s-f-b-nu} cosmological manifestations of the violation of
spin-statistics theorem for neutrinos are considered. In sec.~\ref{s-concl}
we conclude.

\section{Standard cosmological model \label{s-scm}}

As we have already mentioned in the Introduction, the strongest
demand for new physics comes from cosmology. In this connection a
natural questions arise:
how reliable is the standard cosmological model (SCM)?
How much we can trust it? The answers to both questions are
positive. In the following short few-page review we will
advocate this statement. More detailed recent reviews on the basics of 
the modern cosmology can be found in ref.~\cite{cosm-rev}.

The standard cosmological model is very simple and quite robust.
The theoretical setting is the following: \\
1. General Relativity describing gravitational interactions. It is very 
difficult (if possible) to modify this theory at large distances without
breaking some well established fundamental physical principles.\\
2. Assumption (or one can say, observational fact) of homogeneous and 
isotropic distribution of matter in the early universe
in zeroth approximation. This is supported by the smoothness of the
cosmic microwave background radiation with the accuracy better than $10^{-4}$.
Perturbations are treated in the first order analytically 
or numerical stimulated when perturbations rises up to unity.\\ 
3. Knowledge of cosmic particle content and form of their interaction.
Sometimes equation of state is sufficient:
\be 
p_j= f_j(\rho_j)
\label{eqn-of-state}
\ee
with $p_j$ and $\rho_j$ being pressure and energy densities of matter.
In fact there are different forms of matter with different equations 
of state which is indicated by index $j$ here.

The metric describing an isotropic and homogeneous space has the form:
\be
ds^2 = dt^2 - a^2(t)\,f(r) d{\bf r}^2 
\nonumber
\ee
It is a solution of the Einstein equations with
$f(r) = 1+k r^2/4$ and $ k = 0$, $\pm 1$. Theory and observations agree that 
$ k \approx 0$ with a good precision. It means that the spatial geometry 
of our universe is the Euclidean one.

The rate of the cosmological expansion is characterized by function $a(t)$
or more precisely by its logarithmic derivative called the Hubble parameter:
\be
 H = {\dot a}/{a}
\label{H}
\ee
The Hubble parameter, is time dependent, $ H\sim 1/t$, where $t$ is the universe
age. The famous Hubble law is already here:
\be
V= H\, l
\label{Hubble-law}
\ee
where $V$ is the velocity of a distant object (which is not gravitationally binded 
with our Galaxy or local galactic cluster) and $l$ is the distance to it.

The equations which govern the universe expansion, i.e. time dependence
of the  cosmological scale factor $a(t)$ are the following:
\be
\frac{\ddot a}{a} = -\frac{4\pi}{3 m_{Pl}^2}(\rho+3p)
\label{ddot-a}
\ee
where $M_{Pl}=1.2\cdot 10^{19}$ GeV is the Planck mass. The Newton gravitational 
coupling constant is expressed through it as 
$G_N \equiv M^{-2}_{Pl}$. 

Equation (\ref{ddot-a}) looks as the 2nd Newton law describing the
acceleration of a test body induced by matter inside radius $a$. An 
important fact is that
not only mass (energy) creates gravitational force but also pressure.
It allows for accelerated expansion, if $(\rho + 3 p) < 0$. As is usual
in general relativity, for distances $l$ larger than the inverse Hubble
parameter, i.e. for $l>1/H$, the expansion becomes superluminal. These two
effects are necessary for making the universe suitable for our life.

The second equation looks as the conservation of energy of a test body:
\be 
H^2 \equiv \left(\frac{\dot{a}}{a}\right)^2 =
 \frac{8\pi \rho}{3m_{Pl}^2} - \frac{k}{a^2}
\label{H-2}
\ee
If multiplied by $a^2/2$, the l.h.s. is the kinetic energy, while the first term
in r.h.s. is negative potential energy, and the second term is just 
a constant.

The critical (or closure) energy density is defined as:
\be
\rho_c = \frac{3H^2 m_{Pl}^2}{8\pi}
\label{rho-c}
\ee
It is equal to real total energy density for spatially flat universe, that is
for $k=0$.

Measure of energy density of different species $j$ of any gravitating matter
is expressed through the dimensionless parameter:
\be
\Omega_j = \rho_j/\rho_c
\label{Omega-j}
\ee
Evidently if $ k=0$, then $\Omega_{tot}\equiv \sum_j \Omega_j =1$.

A more adequate variable, which is often used instead of time
is the red-shift, $z$:
\be
z+1 = a(t_0)/a(t)
\label{z}
\ee
where $a(t_0)$ is the value of the scale factor at the present time, so
the value of red-shift today is $z=0$. For adiabatic expansion, $z$
is equal to the ratio of cosmic microwave background radiation (CMBR)
temperature at some earlier time with respect to its present day value (see
below). 

There is one more very important equations, though not an independent one,
namely, the covariant energy-momentum conservation, 
\be
D_\nu T^\nu_\mu \equiv  T_{\mu;\nu}^\nu = 0.
\label{DT}
\ee
In our special homogeneous and isotropic case it looks as:
\be
\dot{\rho}+3H(\rho+p)=0.
\label{dot-rho}
\ee
and follows from eqs. (\ref{ddot-a}) and (\ref{H-2}).

General relativity implies automatic conservation of $T_{\mu\nu}$ (\ref{DT}),
due to the Einstein equations:
\be
G_{\mu\nu} \equiv R_{\mu\nu} - \frac{1}{2} g_{\mu\nu} R 
= \frac{8\pi}{M_{Pl}^2} T_{\mu\nu}
\label{ein-eq}
\ee
Indeed the covariant divergence of the l.h.s. identically vanishes and
so must $ T_{\mu;\nu}^\nu $. 
This is a 
result of general covariance, i.e. invariance of physics with respect to 
arbitrary choice of the coordinate frame.

In recent (and not so recent) literature there are some papers where the 
assumption of time dependent ``constants'' is considered, in particular, 
time dependent gravitational coupling constant, $G_N =G_N(t)$ and cosmological
constant $\Lambda (t)$ (about the latter see below, sec.~\ref{s-dark}). 
The authors of these works use there the same standard Einstein 
equations (\ref{ein-eq}) with non-constant
$M_{Pl}$. Enforcing the condition of conservation of 
the Einstein tensor, $G_{\mu\nu}$, and the energy-momentum tensor of matter,
$T_{\mu\nu}$, the authors derive some relations between $G_N (t)$ and
$\Lambda (t)$. However, this procedure is at least questionable. If the 
Einstein equations are derived as usually by functional differentiation of the
total action with respect to metric, $\delta A /\delta g_{\mu\nu} =0$, then
the equation must contain additional terms proportional to (second) derivatives 
of $G_N$ over coordinates. If the least action principle is rejected, then
there is no known way to deduce an expression for the energy-momentum 
tensor of matter.  

The equations of state are usually parametrized as
\be
p = w \rho
\label{p-w-rho}
\ee
In many practically interesting cases parameter $w$ is constant but it
is not necessarily true and it may be a function of time. In this case to 
determine $w(t)$ one has to solve dynamical equation(s) of motion for the
corresponding field(s).

Simple physical systems with $ w= -1, -2/3, -1/3, 0, 1/3, 1$ are known.
They are respectively vacuum state (vacuum energy), collection of
non-interacting plane domain walls and, next, straight cosmic strings,
non-relativistic matter (with $p\ll \rho$), relativistic matter
(with $p=\rho/3$) and the so called maximum rigid equation of state
($p=\rho$). In the last case the speed of sound is equal to the speed of
light (that's why most rigid). It can be realized by a scalar field in 
the course of cosmological contraction. It is strange that matter
with $w=2/3$ is absent in this sequence.

One can see from eq. (\ref{ddot-a}) that if $w<-1/3$ the cosmological
matter anti-gravitates despite positive energy density. Correspondingly 
the universe expands with acceleration. It is worth to note, however, that
if $\rho >0$, any matter in finite region of space has normal attractive
gravity. Only infinitely large pieces of matter may anti-gravitate.

For ``normal `` matter $ \rho > 0$ and $ |\rho| >|p|$, and thus
$ \dot \rho <0$, so energy density drops down in the course of expansion, 
as is
naturally expected. However for vacuum case the energy dominance condition,
$|p|<|\rho|$, is not fulfilled
$ p_{vac}=-\rho_{vac}$ and the vacuum (or vacuum-like) energy density remains
constant despite expansion:
\be
\rho_{vac} = const
\label{rho-vac-const}
\ee

There might be much more strange states of matter,
phantoms, with 
$ w < -1 $~\cite{phantom-2,phantom}. It such a state were realized the
energy density of this kind of matter would
rise in the course of expansion.
As a result of this rise gravitational repulsion would become so strong
that everything 
will be turn apart in the future, not only galaxies and stellar bodies, 
but even atoms and particles. This is the so-called ``phantom''
cosmology. 
In all known to me examples a constant $w<-1$ appears in some
pathological models. However, it is possible that phantom state
could exist only for some finite time, i.e. $w=w(t)$, 
and ultimately the system
returns to good old state with $w\geq -1$.
 
It is instructive to see some simple examples of the expansion 
regime (we present them for the spatially flat case of $k=0$:\\
1. Nonrelativistic matter, $ p=0$:
\be
\rho \sim 1/a^3,\,\,\, a\sim t^{2/3}, \,\, \rho_c = m_{Pl}^2/6\pi t^2.
\label{non-rel}
\ee
2. Relativistic matter, $ p=\rho/3$:
\be
\rho \sim 1/a^4,\,\,\, a\sim t^{1/2}, \,\, \rho_c = 3m_{Pl}^2/32\pi t^2.
\label{rel}
\ee
3. Vacuum(-like), $ p=-\rho$:
\be
\rho = const,\,\,\, a\sim e^{H t }.
\label{vacuum}
\ee
For normal matter the Hubble parameter decreases with time, $ H\sim 1/t$,
while for vacuum $ H= const$. In the case of phantom cosmology $H$ would
reach singularity in finite time and become infinitely large.

Presented above equations are valid of course in the ideal case of
completely homogeneous universe. So they are applicable to the early universe
when the matter was practically homogeneous, or to the late universe on
large scales. 
As we mentioned above, the rise of perturbations and formation of the large
scale structure (LSS) of the universe are treated in the first order approximation
to the Einstein equations when the perturbations are small. Initially this
is true. When perturbations became large, analytic approximations do not work
and numerical simulations are applied. An analysis of LSS formation is an
important part of SCM, see sec.~\ref{s-lss}.
Comparison of the data with theoretical calculations
allows to determine cosmological parameters and to confirm (or reject) the SCM. 
It is probably the weakest part of the construction but since the cosmological
parameters are determined in several independent ways, the impact of theoretical 
ambiguities is strongly diminished.

Let us now briefly describe main epochs in the universe evolution.\\
1. Beginning - unknown. Maybe time did not exist before creation?\\
2. Inflation, i.e. period of fast (exponential) expansion which set up
the frame for creation of our universe. It surely existed. Cosmological 
inflationary  stage is practically an experimental fact. \\
3. Baryogenesis, generation of excess of matter over antimatter. Baryogenesis
must be a dynamical process and not just a result of charge asymmetric
initial conditions. In the latter case inflation would not be possible.\\
4. Thermally equilibrium universe, adiabatically cooled down.
Some phase transitions could occur on the way, when with decreasing
temperature grand unification (GUT), electroweak (EW) or QCD symmetries
became broken. At such phase transitions topological solitons, e.g. 
monopoles, cosmic strings or even domain walls could be 
formed~\cite{top-sol}.
\vspace{0.2cm}\\
After that we come to much better known periods when all underlying
physics is well known both theoretically and experimentally.
\vs{0.2cm}\\
5. Neutrino decoupling. It takes place at low temperature, $ T\sim 1$ MeV
and is governed by the standard weak interactions.\\
6. Big bang nucleosynthesis (BBN). This epoch is rather spread in
temperature, $ T=1-0.07$ MeV and time, $t= 1-200$ sec. BBN is one of the 
cornerstones of SCM. A good agreement of theoretical results for abundances 
of light elements was one of the first arguments in favor of the hot 
universe model.\\
6. Onset of the structure formation. At high temperatures the universe was
dominated by relativistic matter. It is called radiation dominated (RD) stage.
Structure formation in relativistic matter was inhibited and it could start 
only when the red-shifted as $1/a^4$ relativistic matter became subdominant
with respect to the non-relativistic matter, which is red-shifted only as
$1/a^3$. This epoch is called matter dominated (MD) regime. 
The change from RD to MD stage took place at the red-shift
$ z_{eq} \approx 10^4$ or $ T \sim 1$ eV.\\
7. Hydrogen recombination. It takes place at $ z\approx 10^3$ or
$ T \sim 3000\,{\rm K}\sim 0.2$ eV. After hydrogen recombination the 
cosmic plasma became practically neutral, CMBR decoupled from matter,
and cosmic photons propagated freely after that. After recombination
neutral hydrogen was not resisted by the radiation pressure against
forming cosmic structures and baryons started infalling
into already evolved seeds of structures made of dark matter.\\
8. Reionization, at $z=10-20$ (?) Formation of first stars.\\
9. Present time, $ t_U = 12-14$ Gyr.

\section{Universe today \label{s-univ-today}}

\begin{figure}
\begin{center}
\includegraphics[height=0.5\textheight]
{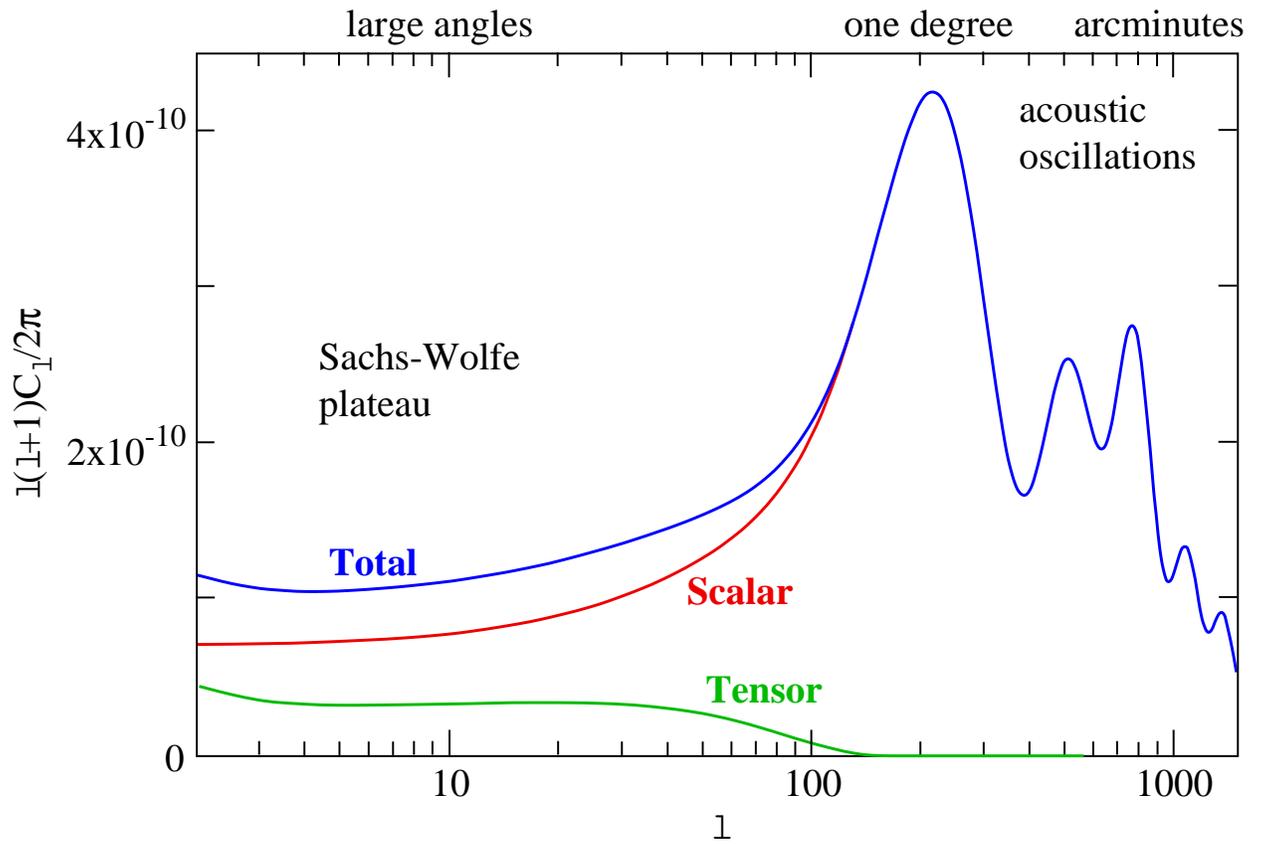}
\caption{ Theoretical angular power spectrum for adiabatic initial
perturbations and typical cosmological parameters. The scalar and tensor 
contributions to the anisotropies are also shown.} 
\label{f-cmbr-teor}
\end{center}
\end{figure}

\begin{figure}
\begin{center}
\includegraphics[height=0.7\textheight]{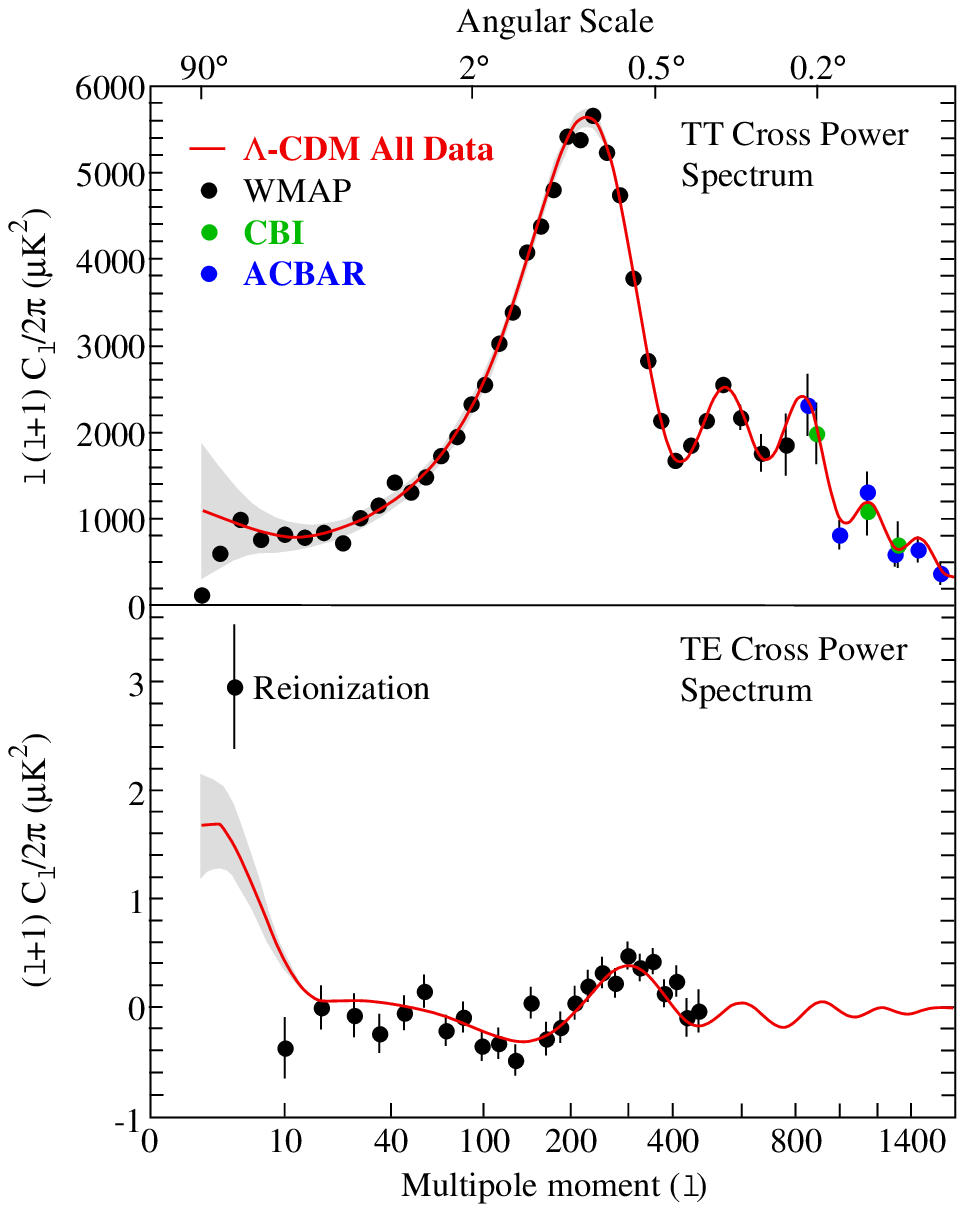}
\caption{Temperature-temperature (TT) and temperature-polarization
TE power spectra. The best fit {${\Lambda}$CDM model} is also shown.
%(Adapted from \cite{26}.)
} \label{f-cmbr-obs}
\end{center}
\end{figure}

Here we will briefly present the main cosmological parameters in the
contemporary universe and comment on the way of their determination.\\
\nin
1. Expansion rate:
\be 
H = 100\,h\,{\rm {\rm km/sec/Mpc}}\,\,{\rm where}\,\,\, h=0.73\pm 0.05
\label{H0}
\ee
The Hubble parameter is determined by direct measurement of the expansion
velocity, i.e. red-shifts of some objects with presumably known luminosity 
(standard candles), as a function of distance and independently by the 
analysis of the angular fluctuations of CMBR.\\
2. Total energy density:
\be
\rho = \rho_c = 1.9\cdot 10^{-29} h^2 {\rm { \frac{g}{cm^3}}}=
10.5\, h^2  {\rm { \frac{keV}{cm^3}}} \approx 10^{-47}\,{\rm { GeV^4}}
\label{rho0}
\ee
It is determined by the known value of $H$ and the fact that the universe
is spatially flat, i.e. $k=0$ (see the following point, 3a).\\
3. Matter inventory.\\
a) $ \Omega_{tot} = 1 \pm 0.1$, from the position of the first peak in
the spectrum of the angular fluctuations of CMBR 
(figs. \ref{f-cmbr-teor}, \ref{f-cmbr-obs})
and analysis of LSS. \\
b) Usual baryonic matter: $ \Omega_{B} = 0.044 \pm 0.004$ 
from independent measurements of the
ratio of heights of the 1st and 2nd CMBR peaks, analysis of BBN, the 
onset of structure formation with small $ \delta T/T$. \\
c) Total dark matter: $ \Omega_{DM} \approx 0.22\pm 0.04 $,
from galactic rotation curves, gravitational lensing, equilibrium
of hot gas in rich galactic clusters, cluster evolution, LSS.\\
d) The remainder:  $ \Omega_{DE} \approx 0.7$. This is the so called
dark energy, a mysterious form of matter which anti-gravitates, i.e.
induces an accelerated expansion. Its value and the fact that it is
antigravitating is found from the dimming of high red-shift supernovae, 
LSS and the universe age.\\
Different pieces of data and their interpretation are 
{\it independent.} All above will be explained in more detail later.
Error bars may be somewhat larger than those presented here.
\vs{0.1cm}\\
There is a mysterious ``cosmic conspiracy'' between different forms
of matter: the energy densities of baryons, unknown non-baryonic dark
matter, and even less known dark energy have comparable values, though
they could differ by many orders of magnitude. This mysterious 
coincidence would be even more pronounced if in addition to the usual
cold dark matter there exist warm dark matter also with $\Omega_{WDM} \sim 0.1$.
An understanding of this conspiracy is a long lasting challenge for 
theorists.

{\it Cosmic microwave background radiation.} This is the famous background of
cosmic photons which was and is one of the strongest arguments in favor
of the big bang cosmology. For a recent detailed review 
see ref.~\cite{cmbr-rev}; a brief digest in~\cite{pdg} is also helpful. 
The energy spectrum of CMBR has a 
perfect equilibrium Planck spectrum, with the temperature
\be 
T = 2.725\pm 0.001\, {\rm { K}},
\label{T-cmbr}
\ee
Correspondingly the number density of such photons in the present day
universe is
$ n_\gamma = 410.4\pm 0.5 $ cm$ ^{-3}$ and their contribution into
the energy density is
$ \Omega_\gamma = (4.9\pm 0.5)\cdot 10^{-5}$.
The temperature of CMBR is almost isotropic over all the sky. Its 
angular fluctuations are very small but their significance is
difficult to overestimate. They allow to observe universe at the very
early stage making a snapshot of the universe at $ z\approx 10^3 $.
The theoretical spectrum of the fluctuations (for some typical values
of the cosmological parameters) is presented in fig. \ref{f-cmbr-teor}.
The coefficients $C_l$ are defined as coefficients in the Legendre 
polynomial expansion of the correlation function of the temperature 
fluctuations:
\be
\langle 
{\frac{\Delta T({\bf n})}{T}\,\frac{\Delta T ({\bf n'})}{T}}
\rangle
= \sum_l\frac{2l+1}{4\pi}\, C_l P_l(\cos \theta),
\label{C-l}
\ee
where ${\bf n}$ and  ${\bf n'}$ are directions of two observations and
$\cos \theta = {\bf n\cdot n'}$.

The comparison of the theory with the measurements permits to 
to measure all basic cosmological parameters, especially if combined with 
other astronomical data (to lift the degeneracy). Observational data are
presented in Fig. \ref{f-cmbr-obs}. (Both figures are taken from
ref.~\cite{cmbr-rev}.)
There is some discrepancy between
the data and theoretical expectations for low multipoles. It can be
explained by the cosmic variance: since the temperature fluctuations
are induced by chaotic density perturbations in the 
early universe, deviations from the average values for a single 
measurement may be large. The measured temperature fluctuations for 
some fixed multipole moment $l$ are averaged over all projections $m$,
$-(2l+1)<m< (2l+1)$. Hence the statistical 
error is small for high multipoles and
may be large for the low multipoles. However, small amplitudes of several low
multipoles indicate to a systematic effect and may deserve a closer attention.
Moreover, all low anomalously small multipoles have the same direction on
the sky, the so called ``evil axis''. Possibly this indicate to some, not
yet understood, physical phenomena.

Now it is a proper place to explain, very briefly, the features of the 
spectrum and their relation to cosmic parameters. Any detailed discussion
will demand a couple more lectures, however. There are two parts in the
spectrum: more or less flat, featureless, at $l\simeq 100$ and an oscillating
part for $l \geq 100$. For smaller $l$, i.e. for longer waves their
length is larger than horizon at recombination. It means that the perturbations
with such large wave length has not yet evolved and their amplitude is equal to
the primordial one. So the part of the spectrum for $l<100$ gives
direct information about spectrum of primordial perturbations. 
The latter is assumed to be a power law type, 
\be
\langle \left( \frac{\delta \rho_k}{\rho}\right)^2\rangle \sim k^n
\label{delta-rho}
\ee
Inflation predicts approximately this type of perturbations 
with $n=1$, i.e. flat or Harrison-Zeldovich spectrum.
Such spectrum corresponds to featureless spectrum of
gravitational potential,
\be
\langle\psi_k^2 \rangle \sim \frac{dk}{k}
\label{psi-k}
\ee
without any dimensional parameter. It can be shown that for $n=1$ the 
power of perturbations of all wave length is the same at the moment
when the corresponding wave length crosses horizon. 
After entering the horizon, if it happened at MD stage,
the amplitudes started to rise because of gravitational instability. All long
waves rise in the same manner and the spectrum is not distorted.
Thus if the universe was at MD stage when long waves (corresponding to  
$l<100$) entered horizon,
we should expect the flat curve for $(\delta T/T)^2$. The curves
in figs. \ref{f-cmbr-teor} and \ref{f-cmbr-obs} are more or
less of this type. However, if after MD stage the universe entered
accelerated regime (we will conventionally call it vacuum dominated, or VD),
the rise of perturbations becomes inhibited. The longer was the wave under
horizon the stronger is the effect. Thus 
$(\delta T/T)^2$ would go down with rising $l$ and there should be a 
minimum between $l=0$ and $l\sim 100$. The observed minimum indeed
indicates to a non-zero vacuum or dark energy and allows to measure
the value the equation of state parameter,
$w \approx -1$. The whole spectrum is well fitted with
$n=1$ and $\Omega_{DE} \approx 0.7$.

The rise at larger $l$ is explained by acoustic oscillations 
at recombination. For shorter wave length and before recombination the
pressure of radiation exceeds gravitational attraction. In other words,
the wave length is shorter than the Jeans wave length. For such a case
perturbations give rise to sound waves. The oscillations of temperature
fluctuations at large $l$ is an imprint of these sounds waves on temperature
of CMBR. 

After recombination when the light pressure on neutral matter
significantly decreased and the Jeans wave length became small,
sound oscillations turned into rising density modes out of which
cosmic structures were subsequently formed. However, before recombination
there are no rising modes, just sound waves. At the moment of horizon
crossing the amplitude of all modes were the same (for flat spectrum of
perturbations). After that each mode was simply red-shifted. Shorter
waves were in such red-shifted regime for a longer time
and their amplitude decreased
more than the amplitude of longer wave. There is one more effect of suppression
of fluctuations at very short wave lengths or large $l>10^3$. This
damping of fluctuations is induced by photon diffusion from high temperature
regions to colder ones and is called diffusion or Silk damping~\cite{silk}.
This is more or less the picture
that we see in the figures. However, on the background of the total decrease
there are some peaks and minima in the spectrum. Their origin is the 
following. Before entering the horizon any mode consists of two
parts, rising and decreasing. Evidently only rising mode survived after 
some time. So in this sense all the modes became coherent standing waves. 
If there were both modes the coherence would be destroyed and the
picture with well pronounced maxima and minima would be destroyed as
well. It can be shown that for adiabatic perturbations only cosine 
mode survives. Thus the sound wave is described by the function
\be
A_k \cos\left[{\bf kx}\right]\,\cos\left[ k \int c_s d\tau \right]
\label{sound}
\ee
where $A_k$ is some slowly varying (known) function related to the spectrum of 
the initial perturbations and
$c_s$ is the speed of sound, for relativistic matter it is 
$1/\sqrt 3$. The amplitude of the mode with fixed $k$ is evidently
\be
A_k \cos\left[ k \int c_s d\tau \right]
\label{sound1}
\ee
The argument under the sign of the integral is just the sound horizon,  
which is slightly 
different from the cosmological horizon (by $\sqrt{3}$). Thus we see that
the maximum amplitude have waves for which $k r_s = n \pi$ with 
$n=1,2,3,...$. (Note that $\tau$ is not the usual physical time but 
conformal time but this is simply technical issue.)

The wave with the largest amplitude is that with $n=1$.
This wave just entered
under horizon at the moment of recombination and did not have time to 
red-shift. This is the largest peak at $l\approx 200$. 

The position of this peak determines 3-dimensional geometry of the
universe. The physical size of the wave length corresponding to this 
largest peak is known. It is $r_s$, the sound horizon at recombination 
(at $T=3000$ K). The value of $l$ where the maximum is reached determines
the angle under which this $r_s$ is observed. The angle in turn
depends upon geometry of the universe. For open universe the angle would be 
smaller, while for closed universe it would be larger. The position at
$l=200$ corresponds to flat Euclidean geometry. That's how CMBR determines
that $\Omega_{tot} =1$.

There are two more, or even maybe three more, observed peaks. The ratio of the
heights of the odd and even peaks determines the amount of baryonic matter
in the universe. 
These peaks corresponds respectively to 
most rarefied and compressed states. Since the oscillations proceeded in
time dependent background, their amplitude 
depends upon the mass density of the oscillating matter. In our case it
is baryonic matter, because dark matter is not coupled to photons and
does not oscillate. By measurement of the ratio of the peak heights the 
most accurate determination of $\Omega_b$ was achieved.

\section{Inflation \label{s-infl}}

Inflation is a period of exponential (or more generally accelerated) 
expansion in the early universe. It was suggested~\cite{guth} 
as a natural explanation of the very peculiar fine-tunings of the
Friedman cosmology. A good introduction and reviews can be found 
in ref.~\cite{infl-rev}. It's interesting that the search and discovery
of inflationary solution was stimulated by a problem of overabundance of
magnetic monopoles~\cite{magn-mon} which was not as important as the
purely cosmological problems.

Usually an approximately exponential inflation is considered:
\be
a(t) \sim \exp [ H t]
\label{a-infl}
\ee
with ${ H \approx const}$.
Such a regime is easy to realize by e.g. a scalar field with slowly
decreasing potential $U(\phi)$. The energy-momentum tensor of
$\phi$ has the form:
\be
T_{\mu\nu} = \partial_\mu \phi \partial_\mu \phi
- (1/2)g_{\mu\nu} \left[\partial_\alpha \phi \partial^\alpha \phi
- U(\phi) \right]
\label{T-scalar}
\ee
In slowly varying potential $\phi \approx const$, 
derivative terms in $T_{\mu\nu}$ can be neglected, thus 
$T_{\mu\nu}\sim  U(\phi) g_{\mu\nu}$ and
$ p\approx -\rho$. This is what is necessary for exponential expansion.

There are two  natural questions: why do we need such a regime and if
inflation is the only known way to create the observed universe. An
argument in favor of positive answers to both questions is a
short list of inflationary achievements which includes at least 
the following:\\
1. Inflation solves the problem of flatness and predicts that universe today
is flat with the accuracy
$ |\Omega -1| < 10^{-4}$ (the number at the r.h.s. is the magnitude of
inhomogeneities on horizon).
Without inflation $\Omega$ must be fine-tuned to unity with
the precision of $ 10^{-15}$ at BBN and
$ 10^{-60}$ at the Planck era. With inflation it goes 
automatically.\\
2. Inflation makes all the observed universe {causally connected.} 
If not inflation, the physically connected regions on the sky would be  
very small, $ < 1^o$, 
while CMBR comes almost the same from all the sky.\\
3. Inflation explains the origin of expansion. This is simply because
of antigravitating character of an almost constant scalar field,
$3p+\rho <0$.\\
4. Inflation makes the universe almost
homogeneous and isotropic at the present-day Hubble volume.\\
5. Inflation creates small inhomogeneities but at astronomically large 
scales, which later became seeds of large scale structure (LSS) formation.
For successful realization of that
the mass of the inflaton should be
$ m_\phi \sim 10^{-5} m_{Pl}$ or it should have an unnaturally weak 
self-coupling, $ \lambda \phi^4$ with
$ \lambda \sim 10^{-14}$.

Generation of primordial density perturbations by a scalar field is
possible because a
scalar field, even with ${ m=0}$, is not conformally invariant.
Life is possible only because of breaking of scale invariance in scalar
field theory.

Inflation not only solves all the above mentioned problems, which
were a nightmare in the old Friedman cosmology, 
in an economic and elegant way but also made an important predictions
of $\Omega =1$ and 
of adiabatic Gaussian density perturbations 
with flat Harrison-Zeldovich spectrum, 

To solve the above mentioned problems, the inflationary period should last 
60-70 Hubble times, i.e. the universe should exponentially expand by the 
factor $e^{60}-e^{70}$. The precise necessary  value depends upon the
(re)heating temperature when inflation ended. When the inflaton field 
$\phi$ approached  zero value of its potential and $H$ became 
smaller than the inflaton mass, $\phi$ started 
to oscillate near the origin and create particles. It is almost as is 
described in the Bible ``Let there be light'' - a dark vacuum-like state 
exploded and created hot universe. This was a moment of 
Big Bang.
(Re)heating temperature is model dependent and most probably is not 
too high,
$ T_{rh}< E_{GUT}\sim 10^{15}$ GeV. In this case no unwanted relics would
be produced, in particular, no magnetic monopoles, which otherwise could
overclose the universe~\cite{magn-mon}. The initial hot universe might be far 
from thermal equilibrium ( see e.g.~\cite{cosm-rev,preheat}) 
and it could facilitate baryogenesis~\cite{infl-bs}.

To conclude inflation is practically 
an experimental fact! It is strongly confirmed by:\\
1. Observed ${ \Omega = 1}$.\\
{2. Flat spectrum of perturbations.}\\
{3. By absence of other way to create our universe.} 
However, be aware of danger of no-go
theorems in physics, e.g. the theorem about impossibility to combine 
internal and space symmetries was overruled by supersymmetry. For 
discussion of competing with inflation cosmological
models see e.g. ref.~\cite{osc-univ}.

To realize inflation, a new field, inflaton, is necessary which is  
absent in the standard model of particle physics.
Competing models also demand new physics and possibly even much more
of it.

\section{Baryogenesis \label{s-baryo}}

According to astronomical observations, the universe,
at least in our neighborhood, is strongly charge asymmetric: it is populated
only with particles, while antiparticles are practically absent. A small
number of the observed antiprotons or positrons in cosmic rays can be
explained by their secondary origin through particle collisions
or maybe by the annihilation of dark matter.
Any macroscopically large antimatter domains or objects (anti-stars,
anti-planets or gaseous clouds of antimatter), if exist, should be
quite rare. As we see in what follows, there are plenty of baryogenesis
scenarios which predict either charge symmetric universe at very large
scales or, even more surprising, an admixture of antimatter
in our vicinity with potentially noticeable amount but still compatible 
with the present observational restrictions.

In connection with the observed asymmetry between matter and antimatter
the first question to address is whether the observed predominance of matter 
over antimatter is dynamical or accidental? The former should be
generated by some physical processes in the early universe starting
from a rather arbitrary initial state, while the latter could be
created by proper initial conditions? Such a question was sensible a quarter
of century ago, but now with established inflationary cosmology, the
answer may be only that any cosmological charge asymmetry could have
been generated dynamically~\cite{ad-baryo-rev,ad-ms-yz}.
The reason for that is the following.
Inflation is incompatible with conserved nonzero baryonic charge density.
Indeed, sufficient inflation, lasting for about $\sim$ 70 Hubble times,
could proceed only if the energy 
density was approximately {\it constant}, see eq. (\ref{H-2}),
and $ a \sim \exp ( Ht)$ if $ H= const$.
If baryons were conserved, the energy density associated
with baryonic charge ({baryonic number})
cannot be constant and inflation could last at most
{4-5 Hubble times.} Indeed any conserved charge evolves in 
the cosmological background as $B\sim 1/a^3 $. Correspondingly
the energy density associated with this charge cannot be constant 
as well and evolves as:
\be
 \rho_B \sim 1/a^n,\,\, n=3-4
\label{rho-B}
\ee
At RD-stage above the QCD phase transition, when quarks were massless,
the baryon energy density was subdominant,
\be
\rho_B/\rho_{tot}  \approx 10^{-10} \approx const
\label{rho-b-rho-tot}
\ee
Let us now travel backward in time to inflationary stage.
At inflation the total energy density is supposed to be constant
$ \rho_{tot}= const$. This can be true if
baryons are not included or negligible. The energy density of baryons
rises on the way back as $\rho_B \sim \exp (-4Ht)$ and
for $ Ht > 6$ the sub-dominant baryons became dominant. Thus inflation
could last at most 6 Hubble times which is by far too short.

Thus initial conditions with an excess of particles over antiparticles
are not compatible with inflation and dynamical
generation of excess of ${ B}$ over ${\bar B}$ is necessary. The 
mechanism for that was suggested by Sakharov~\cite{sakharov}. To this
end three following conditions should be fulfilled:\\
{1. Nonconservation of baryons, theoretically natural; proved to be 
true in MSM and in GUT. Cosmology is an ``experimental'' proof: We exist 
{\it ergo} baryons are not conserved. 
\\
{2. Breaking of C and CP, experimentally established in  particle physics.
\\
{3. Deviation from thermal equilibrium. It is always true in expanding
universe for massive particles or/and in the case of first order phase
transitions in the primeval plasma.

There is a plenty of models of baryogenesis in the literature, 
for reviews see~\cite{ad-baryo-rev,bar-rev}. There is only one number to 
explain, the ratio of baryonic charge density to the number density
of CMBR photons: 
\be 
\beta = \frac{n_B-n_{\bar B} }{n_\gamma } =\frac{ n_B}{n_\gamma }\vert_{today}
\approx 6\cdot 10^{-10}
\label{beta}
\ee
Plethora of scenarios can do that but {all require new physics.}
In principle, baryogenesis is possible in the minimal standard model
of particle physics because it contains all the necessary 
ingredients~\cite{krs}:\\
{1. C and CP are known from experiment to be broken.}\\
{2. Baryons are not conserved} because of chiral anomaly (in nonabelian
theory). The charge non-conservation is achieved by some classical
configuration of the Higgs and gauge boson fields, sphalerons
(see however, criticism in ref.~\cite{taup}).
\\
{3. Thermal equilibrium, would be strongly broken if the electroweak phase
transition was first order.}\\
So far so good. However: \\
{1. Heavy Higgs makes 1st order p.t. improbable}.
Another possible source for deviation from 
equilibrium due to non-zero masses of ${ W}$ and ${ Z}$ is too weak:
\be 
\frac{\delta f}{f_{eq}}\sim \frac{H}{\Gamma}\sim 
\frac{m_W}{\alpha\,m_{Pl}} \sim 10^{-15}
\label{non-equil-W}
\ee
2. {CP violation at high T,}
${ T\sim T_{EW}\sim 100}$ GeV, is tiny:
the amplitude of CP-breaking in MSM is proportional to the product of
{the mixing angles} and to
{the mass differences} of all down and all up quarks:
\be
A_- \sim J
(m_t^2-m_u^2)(m_t^2-m_c^2)(m_c^2-m_u^2)
(m_b^2-m_s^2)(m_b^2-m_d^2)(m_s^2-m_d^2)/ T^{12}
\label{A-}
\ee
where 
\be 
J= \sin \theta_{12}\, \sin \theta_{23}\sin \theta_{31}\, \sin\delta
\label{J}
\ee
is the Jarlskog determinant. At $T\geq 100$ GeV, when sphalerons were in action,
$ A_- \sim 10^{-19}$, which is too small.

It is interesting that it is necessary to have 
three quark families to create CP-violation in MSM
through CP-odd phase in CKM matrix. Thus, if baryogenesis would be
efficient in MSM, there is an 
``explanation'' why we need 3 families. Unfortunately this is not
the case and either 
{CP breaking is different in particle physics and cosmology}
and we do not need three families or some modification of MSM 
would allow to create observable baryon asymmetry with 3 (and only 3)
families.

For example one may avoid these problems and create baryon asymmetry
in EW theory {with 3 families}
if one assumes 
time variation of fundamental constants~\cite{nir} such that
\be
m_{Pl}  \sim  m_{EW},\,\,\, 
m_{q_j} \sim m_{EW}\,\,\, {\rm { and}}\,\,\, 
\delta m_{q_j} \sim m_{q_j}
\label{all-MW}
\ee
But this is {\it very new physics.}
May baryogenesis is an indication of
time varying constants,
TeV gravity, and large higher dimensions?!

Next very popular now scenario of creation of charge asymmetric universe is
{baryo-through-lepto-genesis}. According to it the cosmological 
lepton asymmetry, $L$, could be produced by decays of heavy,
$ m\sim 10^{10}$ GeV, Majorana fermions.
Subsequently L transformed into B by electroweak (sphaleron)
processes~\cite{fuk-yan} (for recent reviews see~\cite{l-rev}).
All three Sakharov's conditions are satisfied:\\
1. {L is naturally nonconserved.}\\
2. {Heavy particles to break thermal equilibrium are present.}\\
3. {Three CP-odd phases of order unity could be there.}\\
The model might successfully explain the origin of the baryon asymmetry but
again demands new physics: 
new heavy particles and new sources of CP violation are necessary.

\section{CP violation in cosmology \label{s-cp}}

There are three possible ways for breaking CP symmetry:\\
1. Standard, explicit by complex constants in Lagrangian.\\
2. Spontaneous~\cite{spont-cp}. This can be realized if there exists 
a complex scalar field with two (or several) 
degenerate vacuum states. The universe as a whole should be
charge symmetric but locally could be asymmetric.
Unfortunately domain walls between vacua with different CP-odd
phases should naturally have a huge energy and thus they are cosmologically
dangerous~\cite{domain}. A natural way to avoid this problem is to make our
domain much larger than the present day horizon. However, it make the
model observationally indistinguishable from the standard one.\\
3. Dynamical or stochastic~\cite{ad-dyn,ad-baryo-rev,brand}. 
This mechanism is similar 
to the spontaneous one but without domain walls. 
It can be also realized by
a complex scalar field displaced from the minimum of its potential 
(e.g. by quantum fluctuations during inflation), and not
yet relaxed to the origin during baryogenesis. However, the
field would certainly vanish to the present time and, if so, 
CP-violation in cosmology would have nothing in common with
the observed CP-violation in particle physics.

An attractive feature of this mechanism is
that a rich universe
structure with isocurvature perturbations and even antimatter domains
may be created. Probably it is the only chance to obtain an observational
information about the mechanism of baryogenesis, because not only
one number $\beta$ (\ref{beta}) is explained but a whole function
$\beta(t,{\bf x})$ with interesting observational
features is predicted.
For detailed discussions see lectures~\cite{cp-cosm}

\section{Problem of vacuum and/or dark energy \label{s-dark}}

\subsection{Definition and history \label{ss-def-hist}}
As is well known the Einstein equations allow an additional term proportional
to the metric tensor:
\be
R_{\mu\nu} -\frac{1}{2} g_{\mu\nu}R - \Lambda\, g_{\mu\nu} = 
8\pi G_N\,T_{\mu\nu} 
\label{eqn-lam}
\ee
This addition was suggested by Einstein himself~\cite{ein-lam} in 1918 
in an (unsuccessful) attempt to obtain stationary solutions of 
equations (\ref{eqn-lam}) in cosmological situation. The idea was
that the gravitational repulsion induced by $\Lambda$ could counterbalance
gravitational attraction of the ordinary matter. The fine-tuning looks
unnatural and what's more the equilibrium is evidently unstable.

This was the beginning a long and still lasting story of $\Lambda$-term. 
Its biography sketch looks as follows:\\
1. Date of birth: 1918.\\
2. Names: Cosmological constant, $\Lambda$-term, Vacuum Energy, 
or, maybe, Dark Energy.\\
3. Father: A. Einstein, who later, 
after the Hubble's discovery of the cosmological expansion,
considered his baby as  ``the biggest 
blunder of my life''. 
Some more quotations worth to add here. Lemaitre: ``greatest discovery, 
deserving to make Einstein's name famous''.
Gamow: ``$\lambda$ raises its nasty head again'' (after indications in the 
60s that quasars are accumulated near $z=2$).\\ 
4. Several times and for a long time $\Lambda$-term 
was assumed dead, probably erroneously. 
It is well alive today, but still not safe - many want to kill it.

This new term is called cosmological constant, because 
the coefficient $\Lambda$ must be a constant independent on space-time
coordinates. Indeed, the Einstein tensor is known to be covariantly
conserved:
\be
D_\mu \left( R^\mu_\nu - \frac{1}{2} g^\mu_\nu R \right) \equiv 0 
\label{DG}
\ee
This property is automatic in metric theories. There is a
simple analogy with electrodynamics. The Maxwell equations has the
form:
\be
\partial_\mu F^{\mu\nu} = 4\pi J^\nu
\label{maxwell}
\ee
Owing to antisymmetry of ${ F^{\mu\nu}}$,
\be
\partial_\mu \partial_\nu F^{\mu\nu} \equiv 0
\label{d-fmunu}
\ee
and the current must be conserved,
\be
 \partial_\mu J^\mu =0.
\label{dJ}
\ee
On the other hand, current is conserved because of $U(1)$-invariance
(gauge invariance) and the theory is self-consistent. 

In the case of general relativity the situation is very similar.
Automatic conservation of the Einstein tensor (\ref{DG})
leads to conservation of the sum:
\be
D^\mu \left[ T^{(m)}_{\mu\nu}/ (8\pi m_{Pl}^2) +\Lambda g_{\mu\nu} \right] = 0
\label{DTLambda}
\ee
The energy-momentum tensor is defined as a functional derivative of the
matter Lagrangian and must be conserved due to invariance of the theory 
with respect to arbitrary change of coordinates. Thus 
\be
D_\mu T_{\nu}^{\mu\,(m)} = 0,
\label{DT1}
\ee
Taking into account that in metric theory the metric tensor is covariantly 
constant
\be
{ D_\mu g^\mu_\nu \equiv 0,}
\label{dg}
\ee
we come to an important conclusion that
the cosmological constant must be {\it constant}:
\be
{{\Lambda = const}}
\label{Lambda-const}
\ee
There are plenty of models in the literature with time dependent 
cosmological constant, $ \Lambda = \Lambda (t)$. It is evident 
from the discussion above that such models 
are not innocent. One needs to introduce additional fields to
satisfy the energy conservation condition or to make more
serious modifications of the theory, e.g. to consider
non-metric theories.
The first attempt to make a time-dependent $\Lambda$, was done 
in 1935 by Bronshtein~\cite{brn}. It was 
strongly criticized by Landau by the reasons presented above.

A new understanding of the nature of the cosmological constant
came from quantum field theory. In its language 
$\Lambda$ is equivalent to vacuum energy density:
\be
T_{\mu\nu}^{(vac)} = \rho_{vac} g_{\mu\nu},\,\,\,
\Lambda = 8\pi \rho_{vac} / m_{Pl}^2.
\label{lam-rho-vac}
\ee
Quantum field theory immediately led to very serious problems.
Theoretically vacuum energy should be huge, $\Lambda \approx \infty$.
Mismatch between theory and upper bounds from cosmology was at the
level 100-50 orders of magnitude, depending upon the type of contribution 
into vacuum energy. As a result the scientific establishment acquired
an eclectic point of view:
\be
\infty = 0
\label{infty0}
\ee
This point of view is shared by many people even now. As Feynman said many years
ago about radiative corrections in quantum electrodynamics:
``Corrections are infinite but small''. However, in contrast to electrodynamics
there are well defined contributions into vacuum energy which are known
to be 50 or more orders of magnitude above its observed value (see below).

From 60s to the end of the Millennium vacuum energy was assumed to be
identically zero.
Only a few physicists treated the problem as an important one, 
starting from Zeldovich~\cite{zel-lam}. A non-negligible amount of 
scientists took a more serious attitude to $\Lambda$ only recently
at the end of  XX century when several independent pieces 
of data were accumulated 
which strongly indicated that empty space is not empty and it
antigravitates inducing accelerated expansion. It is still unknown
what induces this acceleration: vacuum with positive energy density
or some other form of energy with negative pressure, $p<-\rho/3$.
What makes things even more mysterious is a
close proximity of $ \rho_{vac} = const$ to the time varying
energy density of matter $\rho_m \sim 1/a(t)^3$ exactly today. The
origin of this cosmic conspiracy is necessary to understand.

\subsection{Accelerated universe, data \label{ss-accel}} 

The data at the end of 90s which led to a revolutionary change in public 
opinion are the following:\\
1.Universe age crisis.
With $ H\geq 70$ km/sec/Mpc the universe would be too young,
$ t_U < 10$ Gyr, while stellar evolution and nuclear chronology
demand $ t_U \geq 13$ Gyr.\\
2. The fraction of matter density in the universe 
happened to be too low: $\Omega_m = 0.3,$. The result was
obtained by several independent ways:
mass-to-light ratio, gravitational lensing, galactic clusters evolution
(number of clusters for different red-shifts, $z=0$ and $z\sim 1 $).
On the other hand,
inflation predicts $\Omega_{tot} = 1$.
Spectrum of angular fluctuations of CMBR (position of the first peak)
measures $\Omega_{tot} = 1\pm 0.05.$. \\
3. Dimming of high redshift supernovae~\cite{hi-z-sn} directly
indicate that at $z\sim 1$ the universe started to expand with acceleration.
This dimming 
cannot be explained by dust absorption because it was found that the effect is
non-monotonic in $ z$. At larger $ z$ the dimming decreases, as one should 
expect if the effect is induced by vacuum or vacuum-like energy.
Indeed, $\rho_m \sim 1/a^3$, while $\rho_{vac} = const$. Expansion goes
with acceleration when $2\rho_{vac}> \rho_m$. For the measured 
present days values, $\rho_{vac} \approx 2.3 \rho_m$, the 
equilibration between gravitational attraction and repulsion took 
place at $ z\approx 0.7$ in accordance with observations.\\
4. LSS and CMBR well fit theory if $\Omega_v \approx 0.7.$
The theory of LSS formation is not free from assumptions but they
are quite natural and testable. The basic inputs are:
gravitational instability based on general relativity, assumption of
flat spectrum of primordial fluctuations (verified by observation of
low multipoles of the angular spectrum of CMBR), assumption on the type
of dark matter: cold dark matter (non-interactive?), and numerical
simulations when perturbations became large.
 
To conclude: it is established by different independent astronomical
observations that vacuum or vacuum-like energy is non-vanishing. Its
relative contribution into the total cosmological energy density is 
large in cosmological scale:
\be
\Omega_v = 0.7\,\,\,{\rm or}\,\,\, \rv\approx 10^{-47}\,{\rm GeV}^4
\label{omega-vac} 
\ee
and negligible on particle physics scale.
All different data are consistent with
\be
{  \Omega_m = 0.3}\,\,\, {\rm and}\,\,\, \Omega_{tot} =1 
\label{om-tot-om-m}
\ee

\subsection{Evolution of vacuum(-like) energy during cosmic history
\label{ss-evol}}

It may be instructive to consider the role of vacuum energy 
during the universe life-time. 
At inflation $\rv \sim 10^{100}\rho_v^{now} $ and was {\it dominant}.
But it was not real strictly constant vacuum energy but 
{vacuum-like} energy of almost constant scalar field, inflaton. After inflaton
decay to elementary particles universe was dominated by normal, presumably
relativistic matter. If the reheating temperature was higher than the grand
unification scale (GUT) a phase transition (p.t.) from unbroken to broken 
GUT state should take place. At such p.t. the change of vacuum energy
was huge:
\be
\Delta \rv \approx 10^{60}\,{\rm { GeV}^4}
\label{rho-gut}
\ee
Whether $\rho_{vac}$ dominated or not, depended upon the order of p.t. For the
second order $\rv$ might be always subdominant, while for the first order p.t.
it could dominate.

Similar picture took place at electroweak p.t. with
\be
\Delta \rv \approx 10^{8}\,{\rm GeV}^4
\label{rho-ew}
\ee
and at QCD p.t. with
\be
\Delta \rv \approx 10^{-2}\,{\rm { GeV}}^4
\label{rho-qcd}
\ee
There is one important point related to QCD p.t.: the  
magnitudes of vacuum energies of gluon and chiral condensates, which have been
formed at this p.t., are known from experiment!
 
After inflation till almost the present epoch $\rv$ was always
{sub-dominant}, with possible exception for 
short periods before completion of phase transitions. It
started to dominate energy density only recently at
$ z\approx 0.3.$

\subsection{Contributions to vacuum energy \label{ss-contrib}}

According to quantum field theory all fields are represented as an infinite
collection of quantum oscillators, each having energy $\omega/2$ (as
in the usual quantum mechanics). The contributions to $\rv$ from bosonic
and fermionic fields have different signs because the bosonic field operators
commute, while  fermionic ones anti-commute. The energy density of a bosonic
vacuum  fluctuations is
\be
\langle {\cal H}_b \rangle_{vac} = \int \frac{d^3 k}{(2\pi)^3}\,
\frac{\omega_k}{2}
\langle a^\dagger_k a_k + b_k b^\dagger_k \rangle_{vac} 
 =\int \frac{d^3k}{(2\pi)^3}\,\omega_k = \tcred{\infty^4},
\label{h-b}
\nonumber
\ee
while that of fermionic one is
\be
\langle {\cal H}_f \rangle_{vac} = \int \frac{d^3 k}{(2\pi)^3}\,
\frac{\omega_k}{2}
\langle a^\dagger_k a_k - b_k b^\dagger_k \rangle_{vac} 
= \int \frac{d^3k}{(2\pi)^3}\,\omega_k = -\infty^4,
\label{h-f}
\nonumber
\ee
where $\omega = \sqrt{k^2 +m^2}$ and $m$ is the mass of the corresponding 
field.

One immediately sees that if there is an equal number of bosonic and
fermionic fields in nature and their masses are equal (at least pairwise)
then the energy of vacuum fluctuations vanishes. This was noticed by
Zeldovich~\cite{zel-lam} three years before supersymmetry was 
discovered~\cite{susy}.

Indeed if the nature is supersymmetric, the number of bosonic degrees of freedom is
equal to the number of fermionic ones: $N_b = N_f$ and, if SUSY is unbroken,
the masses must be equal as well,
$ m_b = m_f$. In this case indeed $\rv =0$. 
In real world, however, SUSY is broken and soft SUSY breaking, required by
renormalizability of the theory, necessarily leads to
\be
\rv \sim 10^8\,\, {\rm GeV}^4\, \neq 0
\label{rho-susy}
\ee
This could be bad news for supersymmetry but the local realization of
the latter which includes gravity, SUGRA~\cite{sugra}, allows for vanishing
vacuum energy even in the broken phase but at the expense of fantastic 
fine-tuning at the level of about $10^{-120}$, because the natural value
of the vacuum energy in broken SUGRA is at the Planck scale:
\be 
\rv \sim m_{Pl}^4 \geq 10^{76}\,\, {\rm { GeV}^4}
\label{rho-sugra}
\ee

As we have discussed in the previous subsection, in the course of cosmological
expansion the cosmic plasma underwent several phase transitions at which vacuum 
energy changed by the amount much larger than the observed today value:
\be 
\Delta \rv \gg 10^{-47}\,\, {\rm { GeV}^4}
\nonumber
\ee
Especially striking are QCD condensates which are known to be non-vanishing
in the today's vacuum state. QCD is well established and experimentally verified 
science. It  leads to the conclusion that
{vacuum is not empty} but filled with quark and gluon
condensates:
\be 
\langle \bar q q \rangle \neq{ 0}, \,\,\,
 \langle G_{\mu\nu} G^{\mu\nu} \rangle \neq 0
\label{qcd-cond}
\ee
both having non-zero (negative and huge on cosmological scale) vacuum energy:
\be 
\rho_{vac}^{QCD} \approx\, - 10^{45} \rc
\label{rho-qcd1}
\ee

The fact that this condensate must exist can be seen from the estimate of the
proton mass. Proton (or neutron) consisting of three light quarks, each having mass 
of a few MeV, should have mass about 10-15 MeV, instead of 940 MeV. This discrepancy 
is solved by existence of the gluon condensate with negative energy. Inside the
proton the vacuum gluon condensate is destroyed by quarks and the
proton mass is:
\be
m_p = 2m_u + m_d - \rho_{vac}\, l_p^3,
\label{m-proton}
\ee
where $l_p \sim 1/m_\pi$ is the proton size and $m_\pi$ is the pion mass.

However outside proton gluon condensate is non-vanishing and its energy
density is about $10^{45}$ larger than the cosmological energy density.
The big question is
who adds the necessary ``donation'' to make the observed
$ \rv>0$ and what kind of matter is it?

\subsection{Intermediate summary \label{ss-summary}}

We can summarize the state of art as follows:\\
1. Huge contributions to $\rv$ are known but mechanism of their
compensation down to (almost) zero observed value is a mystery.\\
2. Observed today $ \rv $ and $\rho_m$ differ only by factor 2, despite 
of very different evolution with time, $\rv = const$ and $\rho_m \sim 1/t^2$.
Why? \\
3. What is the nature of the antigravitating matter? Is it just vacuum energy
or something different? 

Mostly only problems 2 and 3 are addressed in the literature either by
modification of gravity at large scales or by an introduction of a
new (scalar) field (quintessence) leading to accelerated expansion. 

However, most probably all three problems are strongly coupled and can be solved
only together after an
adjustment of $\rv$ down to $\rc$ is understood.

\subsection{Possible solutions \label{sol}}

The list of suggestions which may in principle lead to a solution of all three
problems summarized in the previous subsection is the following (possibly 
incomplete):\\
1. Subtraction constant. If dark energy is simply vacuum energy, then  
only one number has to be explained and theoretically it may have any value.
This uncertainty reflects the unknown level of zero energy or in other words
an arbitrary value of a subtraction constant. There is of course an enormous 
fine-tuning to fix this value just at $\rc$ but formally one cannot exclude this 
ugly solution. 
In the case that dark energy is not vacuum energy and changes with time, the
subtraction constant idea is aesthetically even less appealing but still is
not excluded.
\\
2. Anthropic principle~\cite{anthrop}. This principle excludes very large values 
of vacuum energy because life would not be possible in such universes. It gives
a justification of a good choice of the subtraction constant. The fine-tuning in
this case drops down from 100 orders to 2-3 orders of magnitude.
Lacking any better suggestion, this might be a feasible solution but
still not especially attractive. It reminded unsatisfactory solutions of 
the problems of the old Friedman cosmology before inflationary idea was 
proposed~\cite{guth}. \\
3. Infrared instability of massless fields (gravitons) in De Sitter 
space-time~\cite{infrared}. Massless particles are quite efficiently
produced by gravitational field in exponentially expanding background.
Their energy density may compensate the original vacuum energy
and slow down the expansion. The mechanism looks natural and quite
promising but unfortunately it seems to be inefficient.\\
4. Dynamical adjustment, analogous to axion solution of the problem of
strong CP violation in QCD. It looks most attractive to me and because 
of that it is discussed below in a separate subsection.\\
5. Drastic modification of existing theory: higher dimensions
breaking of general covariance and Lorentz invariance, a rejection of the 
Lagrange/Hamiltonian principle, ... 
Maybe solution is indeed somewhere on this road but it is difficult to 
say today where is it exactly.

There are dozens reviews on the vacuum energy problem now. An incomplete list
of recent ones are collected in ref.~\cite{lambda-rev}.

\subsection{Dynamical adjustment \label{ss-dyn-adjust}}

The basic idea of dynamical adjustment is extremely simple: it is
assumed that there exists a new field $\Phi$ (scalar or higher spin) 
coupled to gravity in such a way that in De Sitter space-time a
vacuum condensate of $\Phi$ is formed which compensates the 
original $\rv$. This would be a manifestation of the general
Le Ch{\^a}telier principle on cosmological scale: the system always
react in such a way as to diminish an external impact.

The dynamical adjustment was historically first~\cite{ad-82} proposal 
to compensate vacuum energy down to cosmologically acceptable level.
Though no satisfactory realization of the idea is found up to now, it
has very attractive properties. First, vacuum energy is never compensated down 
to zero but only to some remainder which is generally of the order of $\rho_c$.
In this sense dynamical adjustment idea predicted (and not postdicted) existence
of dark energy with the density close to $\rho_c$ and thus solved mentioned
above problems
of compensation of the huge vacuum energy and of existence of non-compensated
remnant with the proper magnitude.
Byproducts of dynamical adjustment have many features of
less ambitious models of modified gravity, e.g. an
explicit breaking of Lorentz invariance, and a
time dependent unstable background with stable
fluctuations over it.

The first attempt to realize dynamical adjustment idea was based on a
non-minimally coupled scalar field~\cite{ad-82}:
\be
\ddot \phi + 3H \dot \phi +U'(\phi, R) = 0
\nonumber
\ee
with e.g. $U = \xi R \phi^2/2$. This equation has unstable, rising with time, 
solutions if $\xi R<0$. Asymptotically $\phi$ rises as
$\phi \sim t$ and the De Sitter exponential expansion turns into Friedman
power law one, but this is not due to compensation of $\rv$ by $\rho_\phi$
because the energy-momentum of the latter is not proportional to the vacuum one:
\be
 T_{\mu\nu} (\phi) \neq F g_{\mu\nu}
\label{t-munu-phi}
\ee
and the change of the regime is achieved due to weakening of
the gravitational coupling:
\be 
G_N \sim 1/t^2 
\label{G-of-t}
\ee
This is surely excluded.  
The models with vector~\cite{ad-vect} or tensor~\cite{phantom-2} 
fields were only slightly better but still did not lead to realistic 
cosmologies.

More recently a  scalar with ``crazy'' coupling to gravity
was considered~\cite{muk-ran,ad-kawa}:
\be
  A= \int d^4 x\sqrt{g} \left[ -\frac{1}{2} (R+2\Lambda)
  + F_1(R) +
 \frac{D_\mu \phi D^\mu \phi} { 2\,R^2}
- U(\phi, R) \right] 
\label{A}
\ee
Equation of motion for $\Phi$ has the form:
\be
  D_\mu\left[ D^\mu \phi\,\left(\frac{1}{R}\right)^2\right]
  + U'(\phi) = 0.
\label{DPhi}
\ee
The second necessary equation is the trace of the Einstein equations. In
particular case with $F_1 = C_1 R^2$ it reads:
\be
  - R + 3 \left( \frac{1}{R}\right)^2 \left( D_\alpha \phi \right)^2
  - 4\left[ U(\phi) +\rho_{vac}\right]
 - 6 D^2  \left[ 2C_1 R - \left(\frac{1}{R}\right)^2
  \frac{\left( D_\alpha \phi\right)^2}{R}\right] =
  T^{\mu}_{\mu}
  \label{trace}
\ee
One can check that the solution of this equations tends to
\be
R\sim \rho_{vac} + U(\phi) = 0
\label{sol-R}
\ee
It has some nice features (``almost realistic''), e.g. $H=1/2t$, 
but is unstable and easily runs away from anything resembling realistic
cosmology.

A desperate attempt~\cite{ad-kawa} 
to improve the model by introduction a non-analytic
kinetic term:
\be 
  \frac{(D \phi)^2}{R^2} \rar - \frac{(D \phi)^2}{R\,|R|}.
 \label{absr}
\ee
was unsuccessful too. 

More general action with scalar field~\cite{ad-kawa}:
\be
A = \int d^4 x \sqrt{-g}
[-{m_{Pl}^2}(R+2\Lambda )/16\pi
+F_1 (R)
+ F_2 (\phi, R) D_\mu \phi D^\mu \phi \nonumber \\
+ F_3 (\phi, R) D_\mu \phi D^\mu R - U(\phi, R) ]
\ee
was not yet explored.
Moreover $R_{\mu\nu}$ and $R_{\mu\nu\alpha\beta}$ can be also included.
\vs{0.2cm}\\
To summarize general features of adjustment mechanism:\\
1. Some compensating agent must exist! \\
2. Quite natural to expect that $\rho_{vac}$ is not completely compensated and
$\Delta \rho \sim \rc$.\\
3. A realistic model is needed, it can indicate what is the value of $w$: 
is it (-1), i.e. the non-compensated remnants of vacuum energy is also vacuum
energy or   $w\neq -1$ and a new strange form of energy lives in the universe?

\section{Big bang nucleosynthesis \label{s-bbn}}

BBN or creation of light elements took place in relatively late universe, when she
was from 1 to 200 sec old and temperatures were between MeV and 60-70 keV. 
Physics is well known at this energies. First, it is the usual low energy
weak interaction leading to neutron-proton transformations: 
\be
n + e^+ \lrar p + \bar\nu_e,  \,\,\,
 n + \nu_e \lrar p + e^-
\label{n-p}
\ee
The rate of these reactions became slow with respect to the cosmological 
expansion rate at  $ T\approx 0.7$ MeV. After that the neutron-to-proton 
ratio remains constant (up to the slow neutron decay) and this
determines the starting value of ${ n/p}$-ratio with which nucleons arrived
to formation of light nuclei. The latter started and proceeded very quickly
almost instantly at $T_{BBN}= 60-70$ keV. The exact value 
of $T_{BBN}$ depends upon
the baryon-to-photon ratio, $\beta$ (\ref{beta}). 
This relation allows to 
determine $\beta$ from BBN, especially from deuterium abundance. Though the
deuterium measurements are quite dispersed, the result is in reasonable 
agreement with $\beta$ determined by CMBR.
At $T=T_{BBN}$ all neutrons quickly formed ${ ^4 He}$ (about
25\% by mass) and a little ${ ^2H}$ (${ 3\times 10^{-5}}$ by number),
${ ^3 He}$
(similar to ${ ^2H}$) and ${ ^7 Li}$ (${ 10^{-9}-10^{-10}}$).
The results span 9 orders of magnitude and are well confirmed by the data.

It is interesting that relatively small variation of the Fermi coupling
constant would dramatically change chemical content of the universe. A slight
increase of $G_F$ would lead to a later freezing of $(n-p)$-transformation
and a smaller number of neutrons. In this case primordial universe till formation
of first stars would be purely hydrogenic. In the opposite case of smaller $G_F$
the universe would be helium dominated. In both these extremes the properties 
of the first stars and their evolution would be completely different from
those in our universe.

%\newpage
\begin{figure}
\begin{center}
\includegraphics[height=0.7\textheight]{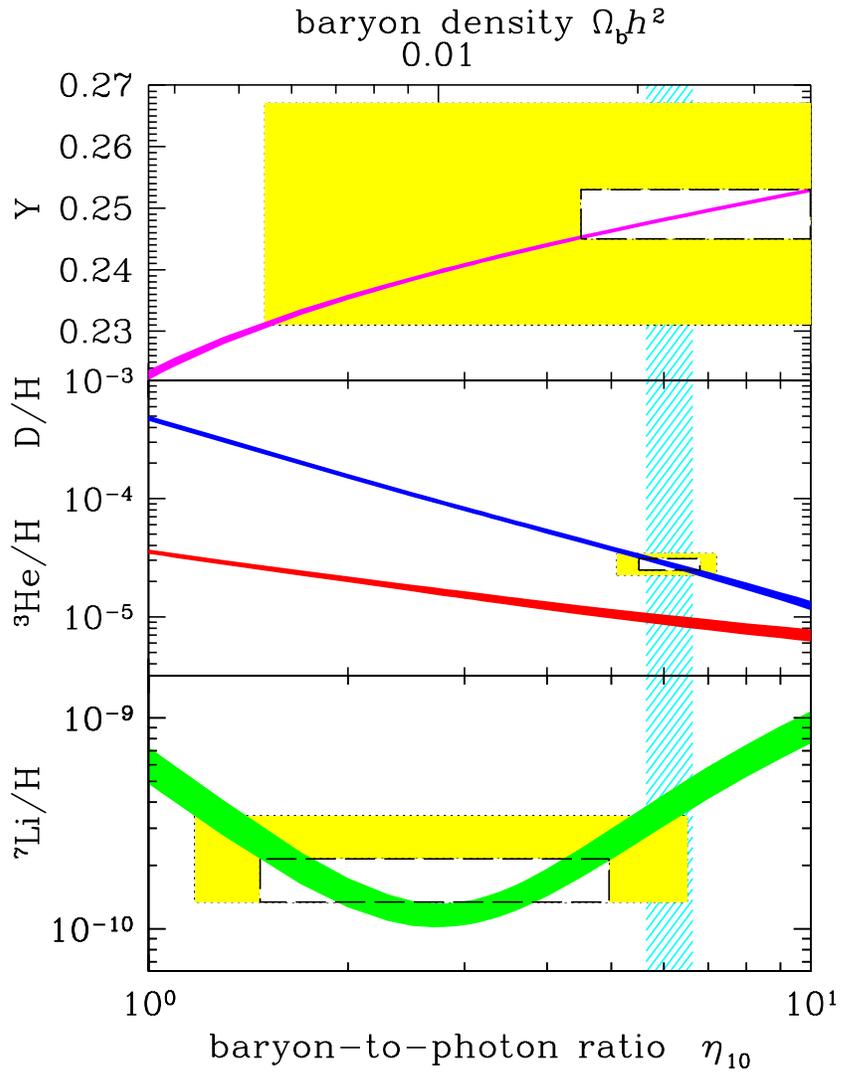}
\caption{
He$^4$, D, He$^3$ and Li$^7$ predicted by the
 standard BBN. Boxes indicate the
  observed light element abundances (smaller boxes: $2\sigma$
  statistical errors; larger boxes: $\pm 2\sigma$ statistical and
  systematic errors). The vertical band is
  the CMB measure of the cosmic baryon density.
} \label{cmb-obs}
\end{center}
\end{figure}
$$$$
\newpage
\begin{figure}[htb]
\begin{center}
\epsfxsize=14cm
\epsfysize=12cm
\epsffile{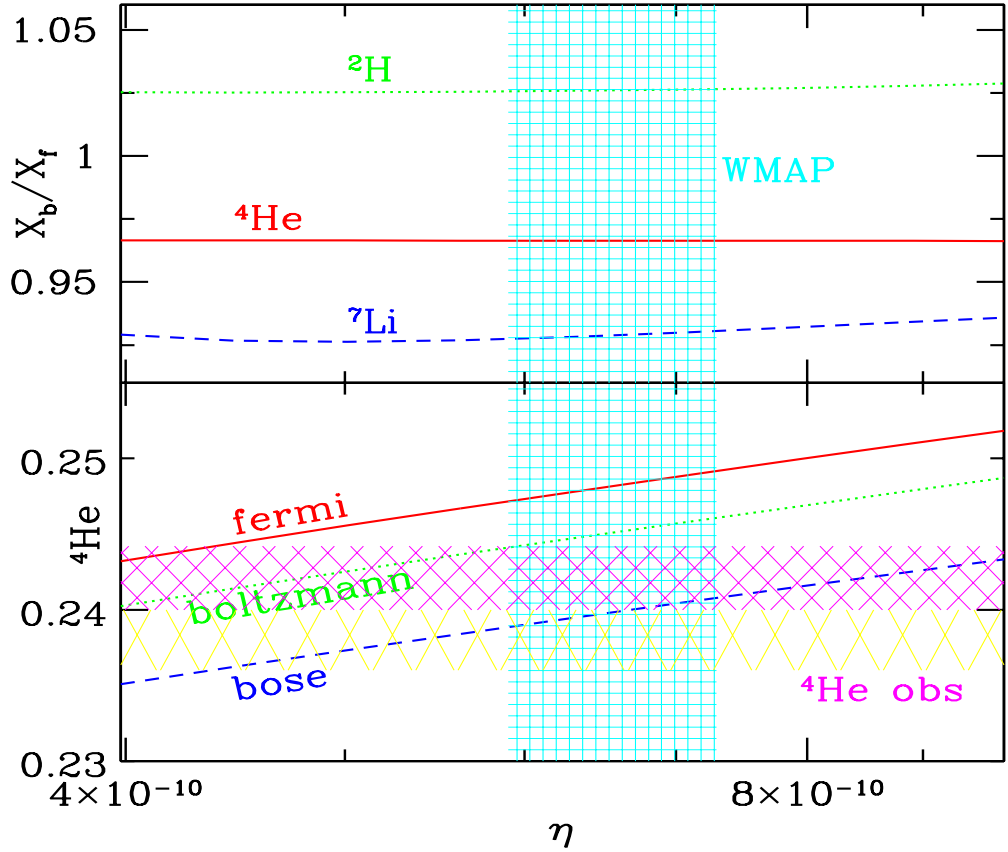}
\end{center}
\caption{Upper panel: the ratios of abundances 
of different elements in the cases of purely
bosonic neutrinos with respect to the standard
fermionic case as functions of
the baryon number density, $\eta$. The vertically hatched (cyan) region 
shows the WMAP $2\sigma$ determination of $\eta$. Lower panel: the
absolute abundance of $^4$He as a function of $\eta$ for the purely
bosonic, Boltzmann, and fermionic neutrino distributions, corresponding
to $\kappa=-1, 0, +1$ respectively. The two skew hatched 
regions show the observation of primordial
helium from ref.~\cite{fieldsolive} (lower, yellow) 
and ref.~\cite{izotovthuan} (upper, magenta), which marginally
overlap at $1\sigma$.}
\label{fig:deut}
\end{figure}
$$$$
\newpage
\begin{figure}
\centerline{\psfig{file=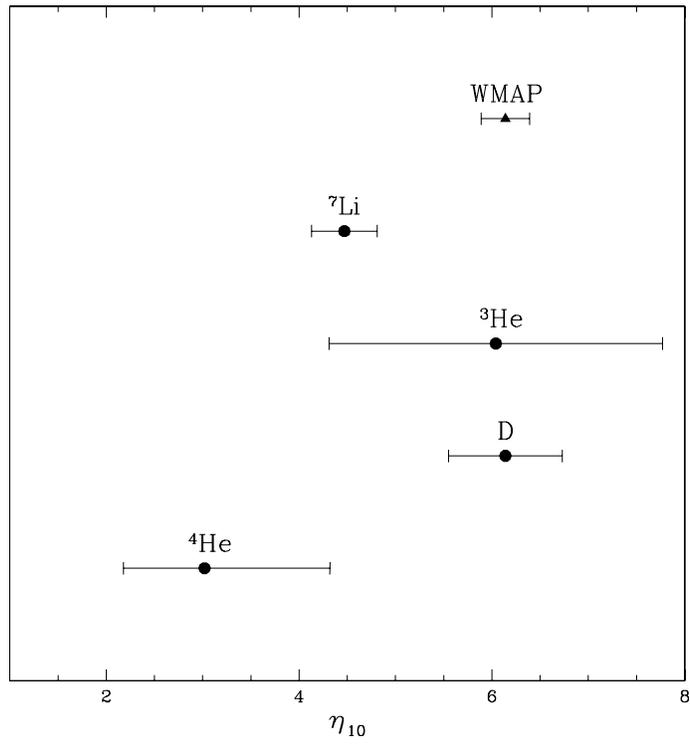,width=10.7cm}}%12cm}}
\vspace*{0pt}%\vspace*{8pt}
\caption{The BBN values for the early universe ($\sim$~20 minutes) 
baryon abundance parameter $\eta_{10}$ inferred from the adopted 
primordial abundances of D, $^3He$, $^4He$, and $^7Li$. 
Also shown is the WMAP-derived value ($\sim$~400 kyr).}
\label{f-eta-bbn}  
\end{figure}   
$$$$
\newpage
As is summarized in review~\cite{pdg} there is a very good agreement between
theory and data on light element abundances with $\beta$ determined form CMBR,
see fig.~\ref{cmb-obs} taken from  ref.~\cite{pdg} 
However, this picture may be too optimistic because there are some conflict 
between $^4He$ and $^2 H$. The former corresponds to lower $\beta$. 
According to ref.~\cite{fieldsolive} $Y=0.238 \pm 0.002$,  
while ref.~\cite{izotovthuan}
finds  $Y = 0.2421 \pm 0.0021$ ($1\sigma$, only statistical error-bars).

These results are shown in fig.~\ref{fig:deut}  as the skew hatched  regions, where 
the upper (magenta) is the results of ref.~\cite{izotovthuan} and the
lower (yellow) is the results of ref.~\cite{fieldsolive}.
In addition, different individual measurements of primordial deuterium demonstrate
large dispersion and indicate to both larger and smaller $\beta$.  
Moreover, the trend for smaller $\beta$ demonstrates $^7Li$. 
For a different from~\cite{pdg} point of view see another review~\cite{bbn-diff}. 
In fig.~\ref{f-eta-bbn}, taken form this work, the baryon-to-photon ratio 
$\eta_{10} = 10^{10}\beta$ is presented as determined from different
light elements abundances and is compared with that found from CMBR. 
There are noticeable deviations.

Most probably the discrepancy will disappear
with future more accurate measurements but if it is confirmed, a strong indication
to new physics will be discovered. 

The data on $^4 He$ and $^2H$ abundances permit to obtain an upper limit
on the number of effective neutrino species, $N_\nu$, participating in BBN, 
because $N_\nu$ effects the cooling rate of the universe and changes the
neutron-proton freezing temperature.
A fair estimate is $N_\nu = 3\pm 1$.

\section{Formation of large scale structure \label{s-lss}}

Formation of galaxies and their clusters is a very essential part of 
the modern cosmology. On one hand, it clearly demonstrates necessity of 
new physics and, on the other, comparison of theory with observations
shows a good agreement of basic principles with the data. For a review
see e.g. ref.~\cite{lss-rev}.

The theoretical input is very simple: rise of the initial small density
perturbations because of gravitational instability. To proceed with 
calculations of the structure formation one needs to know the  
spectrum of primordial fluctuations. Usually it is assumed power law
(\ref{delta-rho}). As we have already mentioned $n=1$ i.e. 
flat, Harrison-Zeldovich spectrum as 
predicted by inflation and confirmed by CMBR at large scales $l \geq 10 Mpc$.

The next important step is an assumption of the
properties of dark matter and dark energy. The former is normally assumed to be 
non-interacting cold dark matter, while dark energy is taken as the vacuum one.
The corresponding cosmology is called $\Lambda$CDM. Interesting possibilities
of self-interacting dark matter, e.g. mirror (for a recent review containing
an impressive list of references see~\cite{mirror}), or warm dark 
matter, remain viable as well.

With these assumption one can make analytical calculations in
linear regime when perturbations are small,
$\delta\rho/\rho\ll 1$. To this end the standard physics, general relativity
and hydrodynamics in gravitational field are used. The results are applicable
to very large scales at which perturbations remain small. These scales
are accessible to CMBR. The angular fluctuations of the latter are in agreement 
with the $\Lambda$CDM picture.

For smaller scales, $l<10$ Mpc,
and larger perturbations, $\delta\rho/\rho\geq 1$, 
when the non-linearity of
equations becomes significant one needs to make
numerical simulations. The latter are of course oversimplified: the mass
of individual particles are taken about $10^6 M_{\odot}$, where 
$M_\odot = 2\cdot 10^{33}$ g is the solar mass, and only gravitational interactions
are taken into account. Still the description happened  to be quite successful
and is confirmed (but strictly speaking not proven) by the data. A large
density perturbations at smaller scales (e.g. isocurvature ones) are not
excluded. Their presence can invalidate the cosmological upper bounds on
neutrino mass recently strengthened in numerous works~\cite{m-nu}
(possibly the list is not complete).

At this stage we can present one more argument in favor of non-baryonic
dark matter. Since protons (and $^4He$-ions) strongly interact with photons,
the fluctuations of baryonic matter can rise only
{after recombination}, i.e in neutral matter. Otherwise photonic pressure 
prevents baryons and electrons from gravitational clumping. 
Density fluctuations at MD regime rise as the first power of the scale factor.
Thus the density contrast in purely baryonic universe may rise at
most by $ 10^3$. In the case of adiabatic fluctuations (proven by CMBR)
\be
\frac{\delta\rho}{\rho} \sim \frac{\delta T}{T} <10^{-4}
\label{delta-rho-delta-T}
\ee
and fluctuations could not reach unity by today in contrast to what we see.
Fluctuations in non-baryonic DM do not interact with light and thus do not
suffer from the pressure of CMBR. Hence fluctuations can rise
at MD-stage prior recombination  starting from redshift $ z\sim 10^4$. 
This allows structures to be formed by the present day.  

Together with the knowledge of $\Omega_{DM} = 0.22$ and $\Omega_B =0.044$
we are forced to conclude that dark matter indeed exists and it is not
the usual baryonic one. Though most of baryons in the universe are invisible
their number is by far smaller than the amount of cosmological dark matter.

As for the possible forms of dark (invisible) matter, there are 
several (too many?) candidates. Among them are the following (but the list
is far from being complete): \\
I. Cold dark matter (CDM):
lightest supersymmetric particle (LSP), $m = (0.1-1)$ TeV;
axion,  ${ m = 10^{-5}}$ eV;
mirror particles;
primordial black holes.\\
II. Warm dark matter (WDM):
{sterile neutrinos ${m =0.1-1}$ keV.}\\
III. Hot dark matter (HDM):
usual neutrinos or light sterile neutrinos; they must be subdominant. Otherwise 
the LSS formation would be inhibited.

Thus new particles which are absent in MSM must exist. It is a striking example
how astronomy predicts an existence of new elementary particles. One can make 
cold dark matter without new stable particles with primordial black holes
with well spread mass spectrum, from a minor fraction of the solar mass up
to millions of solar masses~\cite{ad-js}, but the mechanism of formation
of such black holes also demands new physics.

\section{Unnecessary new physics, or neutrino statistics and cosmology 
\label{s-f-b-nu}}

This is a long standing question if statistical properties of bosons and
fermions may be (slightly) different from the usual Bose-Einstein and
Fermi-Dirac ones.  In fact Pauli and Fermi repeatedly
asked the question if spin-statistics relation could be not exact and 
electrons were not identical but  a little bit different. 

Possible violation of the exclusion principle for the usual matter, i.e. for
electrons and nucleons was discussed in a number of papers at the end of
the 80s~\cite{en-viol}. Efforts to find a  
more general than pure Fermi-Dirac or Bose-Einstein 
statistics~\cite{para} were taken but no satisfactory theoretical 
frameworks had been found. Experimental searches of the
Pauli principle violation for electrons~\cite{exp-viol}  
and nucleons~\cite{exp-bar} have also given negative results.  

However, there is still a lot of experimental freedom to break
Fermi statistics for neutrinos. 
In the case that spin-statistics relation is broken, while otherwise
remaining in the frameworks of the traditional quantum field theory, then
immediately several deep theoretical problems would emerge:\\
1) non-locality;\\
2) faster-than-light signals;\\
3) non-positive energy density and possibly unstable vacuum;\\
4) maybe breaking of unitarity;\\
5) broken CPT and Lorentz invariance.\\
To summarize there is no known self-consistent formulation of a theory
with broken spin-statistics theorem. 
So let us postpone discussion of (non-existing) theory and consider
cosmological effects of neutrinos obeying Bose or mixed statistics~\cite{ad-as}.

If neutrinos were Bose particles 
they could form cosmological Bose condensate:
\be 
f_{\nu_b} = \frac{1}{\exp [(E-\mu_\nu)/T -1] } + C \delta (k),
\label{f-nu-b}
\ee
and make both cold and hot dark matter.
So instead of new particles obeying old physics dark matter could be 
created from the old known particles but with very new physics. More details
can be found in ref.~\cite{ad-as}.

A deviation from Fermi statistics would be observable in 
BBN~\cite{nu-b-bbn,ad-sh-as}. There are two effects resulting in a
change of primordial abundances:\\
1) the energy density of bosonic neutrinos would be 8/7 of normal fermionic $\nu$,
resulting in a rise of effective neutrino species by
$\Delta N_\nu = 3/7$;\\
2) a larger density of $\nu_e$ would lead to smaller temperature of
$n/p$-freezing, which is equivalent to a decrease of $N_\nu$.\\
The net result is that the effective number of neutrinos at BBN 
would be smaller than three: $ N_\nu^{(eff)} =  2.43$. 

We assume, following ref.~\cite{ad-sh-as} that the equilibrium distribution for 
mixed statistics has the form:
\be
f^{(eq)}_\nu = \left[ \exp (E/T) + \kappa \right]^{-1},
\label{f-mixed}
\ee                            
where $\kappa$ interpolates between the usual Fermi statistics, $\kappa = 1$,
to pure Bose statistics $\kappa = -1$. Seemingly, as one can see from
fig.~\ref{fig:deut} the data are noticeably better described by bosonic 
neutrinos. However, observational uncertainties are too large to make a
definite conclusion.

If, together with spin-statistics theorem, CPT and/or unitarity is/are
also broken, the usual equilibrium distributions
would be distorted too and the effects can be accumulated with time
and become large.

\section{Some more unnecessary physics \label{s-more}}

Unfortunately time and space bounds do not allow  
for any detailed discussion of 
other unnecessary new physics as e.g. breaking of Lorenz invariance, 
violation of the CPT theorem, 
possibility of (large) higher dimensions, etc. The situation with all that
is unclear and it could be that the related phenomena are not realized
Though the probability of existence is low but stakes are high and new
physics without any restrictions, except for experimental ones, 
could be very interesting. Probably cosmos will be
the best place to look for these completely news effects. 

A possibility of astronomical manifestations of higher dimensions, $D>4$,
looks exciting. It opens practically unrestricted
room for theoretical imagination. However, it may be difficult to
prove their existence. We live in 4-dimensional 
world and most probably will not be able to penetrate other dimensions.
It may be a non-trivial task to distinguish between
phenomena which came to us from higher D or have a simpler (or even more
complicated) explanation in $D=4$. 

A very interesting effect which may arise from higher dimensions is electric
charge nonconservation~\cite{el-rub}. However, electric charge might be
visibly nonconserved if photons are a little massive and in this case
black holes could capture charge particles without electric hairs 
left outside~\cite{el-bh}. Maybe ``evil axis'' could be explained by an
electric charge asymmetry of the universe?

Probably in 4D world any energy nonconservation is absolutely forbidden
and if it were observed, this would be an unambiguous indication that we 
are connected to $D>4$ space. 

\section{Conclusion \label{s-concl}}

Cosmology unambiguously proves that there is new physics (far) beyond the 
minimal, and not only minimal, standard model of elementary particle world.
As we have seen, inflation, which is practically an experimental fact
demands a new fields or fields which had induced exponential expansion, creating
``correct'' universe. These field have to be very weakly coupled to the ordinary
fields/particles and are absent in MSM.

Baryogenesis is impossible in MSM as well. Either heavier fields  or
time variation of fundamental constants are necessary. In particular the 
amplitude of CP-violation might be much larger in the early universe
than in particle interactions at the present time.

It is proven that the bulk of matter in the universe is not the ordinary 
baryons. There are several good candidates for dark matter particles 
but it is still unknown which one is in reality. This is a primary 
problem for experiment and/or astronomical analysis. 
There is a possibility to make cosmological dark matter 
out of light massive neutrinos but at the expense of breaking spin-statistics
theorem. The price is very high but the consequences would be exciting.

The 70\%-bulk
of matter in the universe, dark energy, is even more weird than 'simple'' 
dark matter. In contrast to 
the ordinary matter, this dark energy induces gravitational 
repulsion. There can be a simple phenomenological description of cosmological
acceleration by a tiny vacuum energy or by a very light scalar field
(do we need that field if vacuum energy can do the job?), but
only if one forgets about formidable problem of huge discrepancy, by 100-50 orders
of magnitude, between the natural value of vacuum energy and the astronomically 
measured value. If one recalls that there are experimentally known huge
contributions into vacuum energy from the quark and gluon condensates, the
problem of vacuum energy compensation becomes even more striking.

One should also remember about cosmic conspiracy of the same magnitude of
different forms of matter: baryonic, non-baryonic, and dark energy. Maybe a
solution of this problem will indicate to some new physical effects and
help to solve the vacuum energy problem?

Speaking about more revolutionary ideas of breaking
unitarity, spin-statistics theorem, CPT, least action principle, etc
one should be aware of Pandora box of consequences if sacred principles
are destroyed. To this end a quotation from ``Karamazov brothers'' by
Dostoevsky: ``If God does not exist anything is allowed.'', may be 
appropriate.

\end{document}